\documentstyle[11pt,newpasp,twoside,epsf]{article}

\begin{document}

\title{CS J$= 7 \rightarrow 6$ Mapping of Massive Star Formation Regions 
       Associated with Water Masers}
\author{C. Knez, Y. L. Shirley, N. J. Evans II, \& K. E. Mueller}
\affil{Department of Astronomy, The University of Texas at Austin,
       Austin, Texas 78712--1083}

\begin{abstract}24 cores have been mapped in CS J=7{$\rightarrow$}6
at the CSO.  From the spectra we determine core sizes and 
virial masses.  Combining results from the CS and dust continuum 
studies for M8E, we use Monte Carlo simulations for the 
CS emission to get radial profiles.

\end{abstract}

\section{Introduction}
In order to sample cores in early stages of evolution, Plume et al. (1992, 1997) 
began a CS survey of H$_2$O masers associated with star formation regions.  H$_2$O 
masers indicate high-density regions (n $>$ 10$^{10}$ cm$^{-3}$) and are believed 
to be in an earlier stage of evolution than ultracompact HII regions.  Because the CS 
J=7{$\rightarrow$}6 line has a high critical density (n$_{crit} \approx 2.8 $x$ 10^7 
$ cm$^{-3}$), we can probe the denser parts of the star forming cores.  Shirley et al. 
(2001) have mapped the CS J=5{$\rightarrow$}4 transition (n$_{crit} \approx 8.9 $x$ 10^6 
$ cm$^{-3}$) which probes slightly less dense gas than the CS J=7{$\rightarrow$}6 line.  
From the CS spectra, we determine core sizes, virial masses, CS intensities, and radial
profiles.  Using Monte Carlo simulations for CS emission with the temperatures for 
given density profiles from the 1-D dust continuum models (Mueller et al. 2001), 
we can compare the spatial extent of the gas and the dust in the star forming regions.

\section{Results from Data}

The size of each core was calculated by deconvolving the beam size from the size of 
the half power contour integrated intensity.  Sizes  were 
determined; however, 7 of those sources were not included in the statistics because 
their deconvolved sizes were less than half of the beam size.  About 50$\%$ of 
the cores had deconvolved sizes greater or comparable to the CSO beam which is 
consistent with a power-law density distribution (Mueller et al. 2001).  The virial
mass contained within the calculated size was also determined.  The CS 
J=7{$\rightarrow$}6 survey is biased against the highest mass cores observed 
in CS J=5{$\rightarrow$}4 due to observational exigencies (i.e. 
they have not been mapped).  The average 
size of the cores was 0.27 $\pm$0.14 pc.  In order to compare the virial 
masses of the CS5-4 and CS7-6 studies, we used the CS5-4 core sizes 
as our standard (Shirley et al. 2001).  The average ratio of the CS7-6 
to CS5-4 virial masses is 0.72.  This result shows that the emission of 
CS7-6 is more compact than the CS5-4 emission.

\section{Modeling of M8E}
 
The contour map of M8E is roughly symmetrical thus making this source a good
candidate for 1-D modeling. The density and temperature profiles were determined
from dust models (Mueller et al. 2001).  The best fit power law to the dust 
data was p=1.75.  Comparing the density and temperature profiles to the 
intensity profile for the CS data, we find that the gas seems to prefer a value
of p between 1.5 and 1.75 (see Figure 1).  This slight disagreement between
the dust and the gas can be due to the assumptions in both the dust and gas 
modeling.  For example, we assume a constant molecular abundance but 
in reality the abundance may vary with radius.

\begin{figure}
\plottwo{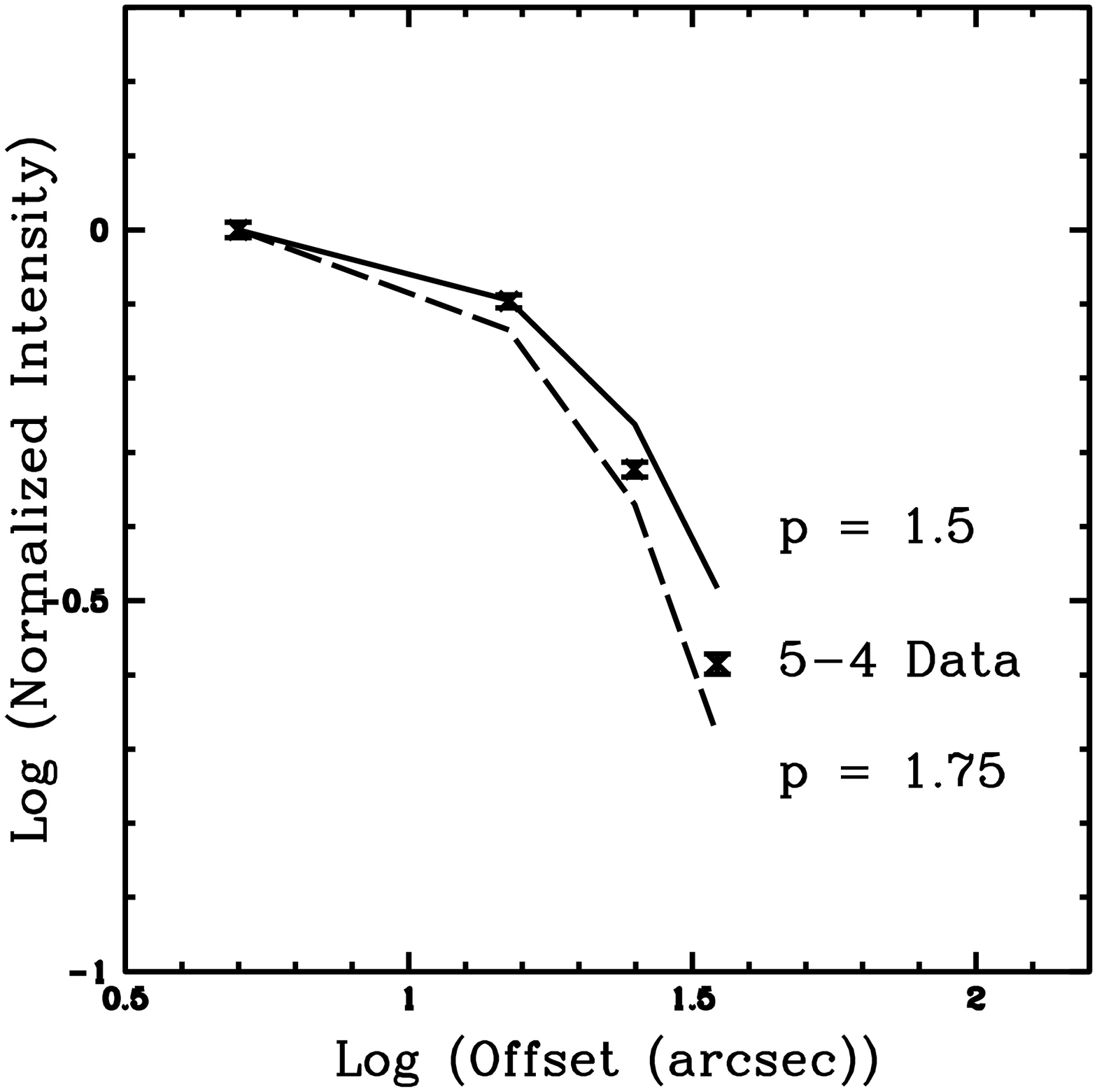}{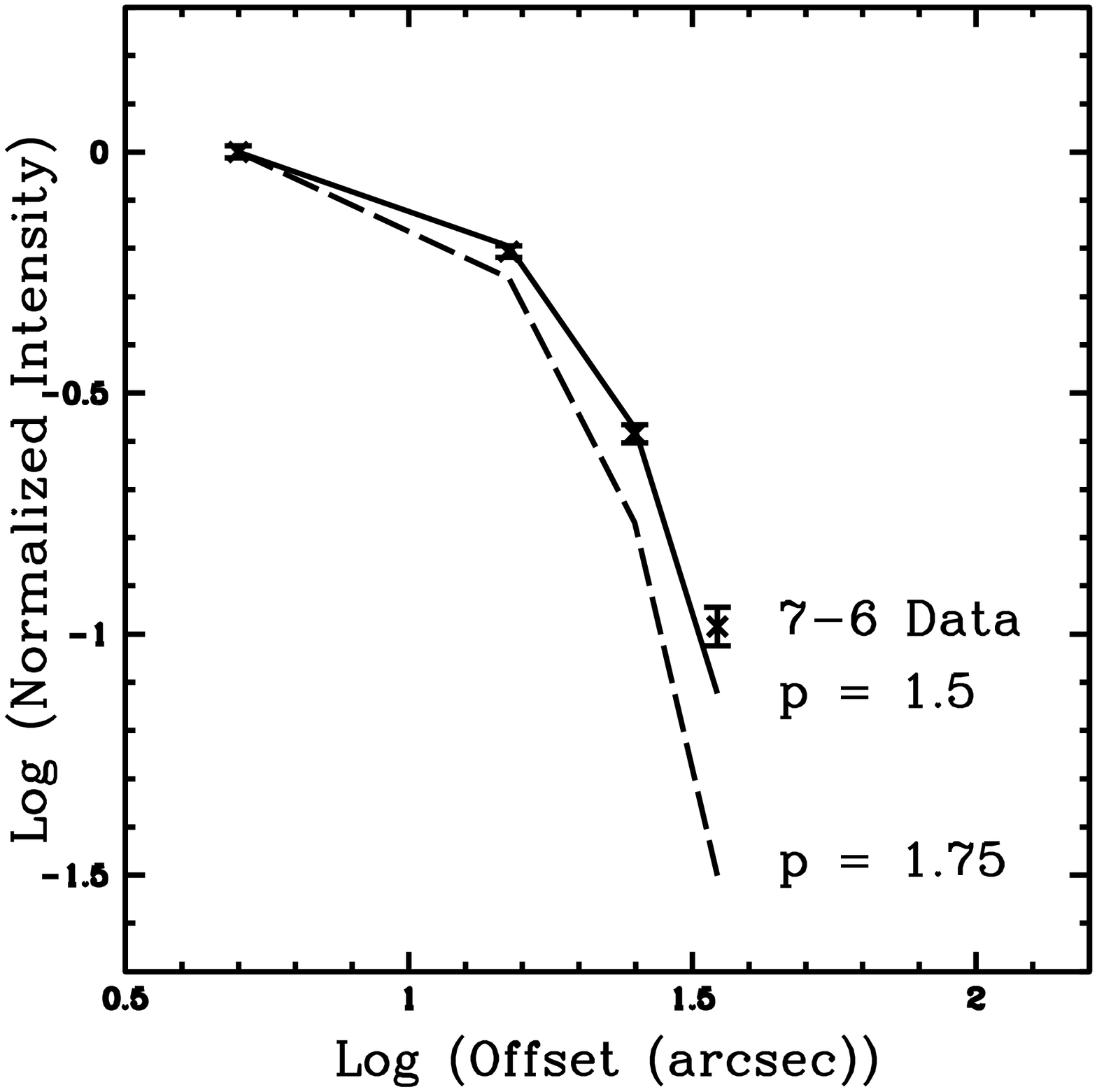}
\caption{Radial profiles of the intensity of CS5-4 (left) and CS7-6 
(right) from the data and the model.  The solid and dashed lines 
represent models with p=1.5 and p=1.75, respectively.}
\end{figure}


\begin{references}

\reference Mueller, K. E., Shirley, Y. L., \& Evans, N. J. II, 2001, these 
	   proceedings
\reference Plume R., Jaffe, D. T., Evans, N. J. 1992, ApJ Supp, 78, 505
\reference Plume R., Jaffe, D. T., Evans, N. J., Mart{\'{\i}}n-Pintado, J., 
           \& G{\'{o}}mez-Gonz{\'{a}}lez, J. 1997, ApJ, 476, 730
\reference Shirley, Y.L., Evans, N. J. II, Mueller, K. E., Knez, C. \& 
	   Jaffe, D. T., 2001, these proceedings


\end{references}
\end{document}